\documentclass{aa}
\usepackage{txfonts}
\usepackage{epsfig}

\def\Swift{\emph{Swift}}
\def\etal{et al.\ }
\def\til{\ensuremath{\sim\,}}
\def\sqiglt{\hbox{\rlap{\lower.55ex \hbox {$\sim$}}\kern-.05em \raise.4ex \hbox{$<$}\,}}
\def\sqiggt{\hbox{\rlap{\lower.55ex \hbox {$\sim$}}\kern-.05em \raise.4ex \hbox{$>$}\,}}

\def\tn{\ensuremath{T_{90}}}

\def\chisq{\ensuremath{\chi^2}}
\def\cms{\ensuremath{$cm$^{-2}}}
\def\G{\ensuremath{\Gamma}}

\begin{document}


\title{The \Swift\ Burst Analyser I: BAT and XRT spectral and flux
evolution of Gamma Ray Bursts}

\author{P.A. Evans\inst{1}\thanks{pae9@star.le.ac.uk} \and R.Willingale\inst{1}
\and J.P. Osborne\inst{1} \and P.T. O'Brien\inst{1} \and K.L.
Page\inst{1} \and C.B. Markwardt\inst{2,3,4} \and S.D. Barthelmy\inst{2}
\and A.P. Beardmore\inst{1} \and D.N. Burrows\inst{5} \and C.
Pagani\inst{1} \and R.L.C. Starling\inst{1} \and N. Gehrels\inst{2}
\and P. Romano\inst{6}}

\institute{Department of Physics and Astronomy, University of Leicester,
Leicester, LE1 7RH, UK \and NASA/Goddard Space Flight Center,
Astrophysics Science Division, Greenbelt, MD 20771, USA \and 
CRESST/Center for Research and Exploration in Space Science and
Technology, 10211 Wincopin Circle, Suite 500, Columbia, MD 21044, USA
\and Department of Astronomy, University of Maryland College Park, College Park, MD 20742, USA 
\and Department of Astronomy and Astrophysics, 525 Davey
Lab., Pennsylvania State University, University Park, PA 16802, USA
\and INAF, Istituto di Astrofisica Spaziale e Fisica Cosmica, Via U La
Malfa 153, I-90146 Palermo, Italy
}

\date{Received / Accepted}

\abstract
{Gamma Ray Burst models predict the broadband spectral evolution and the
temporal evolution of the energy flux. In contrast, standard data analysis tools and
data repositories provide count-rate data, or use single flux
conversion factors for all of the data, neglecting spectral evolution.}
{To produce \Swift\ BAT and XRT light curves in flux units, where the
spectral evolution is accounted for.}
{We have developed software to use the hardness ratio information to
track spectral evolution of GRBs, and thus to convert the count-rate
light curves from the BAT and XRT instruments on \Swift\ into accurate,
evolution-aware flux light curves.}
{The \Swift\ Burst Analyser
website\thanks{http://www.swift.ac.uk/burst\_analyser} contains BAT, XRT
and combined BAT-XRT flux light curves in three energy regimes for all
GRBs observed by the \Swift\ satellite. These light curves are
automatically built and updated when data become available, are
presented in graphical and plain-text format, and are available for
download and use in research.}
{}


\maketitle

\section{Introduction}
\label{sec:intro}

Gamma Ray Bursts are the most powerful explosions known, and the \Swift\
satellite (Gehrels \etal2004) has revolutionised our understanding of
them, both in filling some gaps in our knowledge and raising new
questions and challenges to theory. See Zhang (2007) for a recent review
of GRBs and the advances made by \Swift.

One of the difficulties inherent in confronting theory with the wealth
of data that \Swift\ has produced is that models predict how the flux
and spectrum of a GRB or its afterglow will evolve, whereas the data are
in units of count rate over some bandpass; in the presence of spectral
evolution the count-rate cannot be seen as a proxy for the flux. 

It is thus desirable to create GRB flux light curves\footnote{We create
light curves in units both of flux and flux density; however for
concision, we use the collective phrase, `flux light curves' to refer to
both types.} which employ a time-dependent flux conversion factor to
account for spectral evolution. Also, since the bandpasses of the
\Swift\ Burst Alert Telescope (BAT; Barthelmy \etal 2005; bandpass:
15--350 keV) and X-ray Telescope (XRT; Burrows \etal2005; bandpass:
0.3--10 keV), are close to each other, it is often informative to
consider the two instruments' data together, extrapolated to a single
bandpass, be it the XRT band (e.g. O'Brien \etal2006; Willingale
\etal2007) or part of the BAT bandpass (e.g. 15--25 keV, Sakamoto
\etal2007).

\begin{figure}
\resizebox{\hsize}{!}{\includegraphics[angle=-90,width=8.2cm]{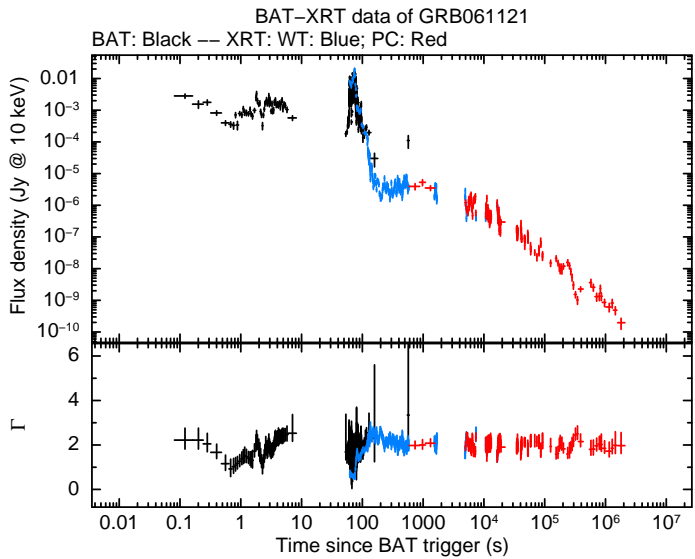}}
\caption{An example of a flux density light curve from the Burst Analyser. The
spectrally-evolving BAT-XRT flux light curve of GRB 061121, is
shown as unabsorbed flux density at 10 keV; the lower panel shows the
evolution of the photon index of the power-law spectrum. The last 3 BAT
data points suffer from a poorly-constrained spectrum; see
Section~\ref{sec:errors} for details.}
\label{fig:example}
\end{figure}

\begin{figure}
\resizebox{\hsize}{!}{\includegraphics[angle=-90,width=8.2cm]{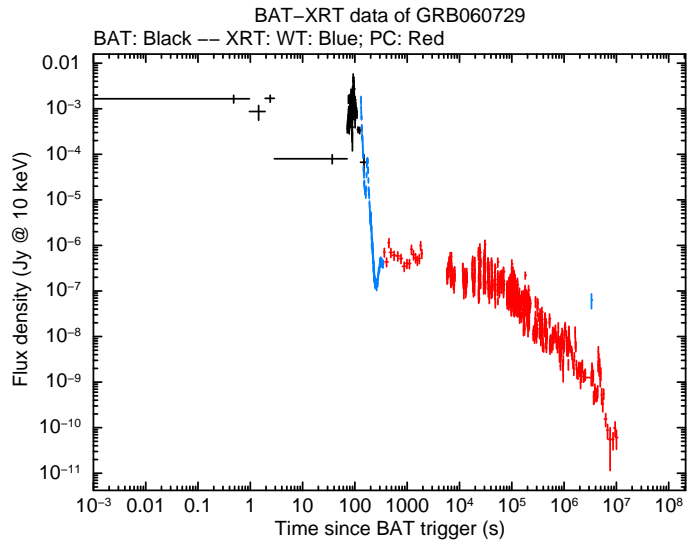}}
\caption{The flux density light curve of GRB 060729 from the Burst
Analyser. Accounting for spectral evolution shows the flux decline
during the steep decay to be more rapid than in count space, and reveals
the turn-on of the afterglow.}
\label{fig:paper2}
\end{figure}

We have therefore created the \Swift\ Burst Analyser. In this first
paper relating to the facility, we present an online repository of BAT
and XRT unabsorbed flux light curves in three energy regimes: 0.3--10
keV flux, 15--50 keV flux, and the flux density at 10 keV.  The Burst
Analyser provides BAT and XRT flux light curves separately and combined;
an example is shown in Fig.~\ref{fig:example}. It also includes a time
evolution history for each instrument of the counts-to-flux conversion
factor, and of the spectral photon index, \G\ (i.e.\ for a power-law
spectrum with the number of photons at energy $E$ is given by $N(E) dE
\propto E^{-\Gamma}$). In addition, we provide BAT flux light curves
where spectral evolution is not included, for comparison with the
non-evolving XRT data already available
online\footnote{http://www.swift.ac.uk/xrt\_curves\label{foot:lc}}
(Evans \etal2007, 2009). One example of the advantage of considering
spectral evolution is shown in Fig.~\ref{fig:paper2}; for GRB 060729 the
rapid decay phase is steeper when viewed in flux space with spectral
evolution accounted for, which allows us to see the turn-on of the
afterglow. (Note that Grupe \etal2007, 2010 also accounted for spectral
evolution in their analysis of this GRB. The evolution-induced dip
feature shown in Fig.~\ref{fig:paper2} is less prominant in the 0.3--10
keV band which is why it is not seen in their analysis.) We consider the
physical interpretation of GRB light curves enabled by our new method in
Paper II (Evans \etal in prep.).

All of the data created by the Burst Analyser are available from:

\begin{center}{\bf http://www.swift.ac.uk/burst\_analyser}\end{center}

\noindent This includes graphical plots and the data in plain-text
format.

In this paper we introduce the \Swift\ Burst Analyser. In
Section~\ref{sec:create} we explain how the light curves are created and
spectral evolution accounted for. Section~\ref{sec:caveats} highlights
the limitations of our method and some recommended checks users should
apply. We also give details of when the light curves are created and how
they can be accessed (Section~\ref{sec:avail}), and the usage policy
(Section~\ref{sec:usage}). 

\section{Creation procedure}
\label{sec:create}

The process of creating the light curves comprises three phases:
generating count-rate light curves and hardness ratios, determining the
counts-to-flux conversion factors, and converting the count-rate light
curves into flux units. Throughout the descriptions of thes phases, any
{\sc ftools} used were called with the default parameters unless
explicitly stated otherwise.

\subsection{Count-rate light curves and hardness ratios}
\label{sec:cr}

The XRT count-rate light curves and hardness ratios used by the Burst
Analyser are taken from the Swift XRT Light Curve
Repository$^{\ref{foot:lc}}$, and their
creation has been documented in detail by Evans et al. (2007, 2009) and
will not be repeated here.

\begin{figure}
\resizebox{\hsize}{!}{\includegraphics[width=8.2cm]{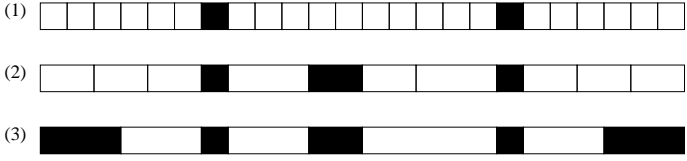}}
\caption{A demonstration of the SNR binning algorithm. In the first pass
(1) the bins marked in black were above the SNR threshold. For the
second pass (2) the sub-threshold input bins were grouped in twos (any
ungrouped input bins were appended to the group they follow, so a few
groups contain three input bins). One of these groups was found to be
above threshold. In the third pass (3) the input bins were grouped in
threes (again, with left-over bins appended if necessary) and two of
these were above threshold. This continues until all of the input bins
are above threshold (i.e.\ the whole plot is black) or the input bins
are grouped into new bins of 40-s or longer, at which point they are all
added to the output light curve.}
\label{fig:binning}
\end{figure}

For BAT, the {\sc batgrbproduct} script (part of the \Swift\ 
software\footnote{http://swift.gsfc.nasa.gov/docs/software/lheasoft/download.html})
is first executed. Among the products created by this script are a
measure of \tn\ and the time range over which it was determined; a
spectrum corresponding to this interval (hereafter, `the \tn\
spectrum'); and an event list of BAT data extending to 2000 s either
side of the trigger. From this event list a 4-ms binned 15--150 keV
light curve is extracted using the {\sc batbinevt} tool, supplied with
relevant detector mask file created by {\sc batgrbproduct}.
Signal-to-noise-ratio (SNR)-binned light curves are built from this 4-ms
light curve using a custom script. The algorithm of this script is
described below and illustrated in Fig.~\ref{fig:binning};  in essence
it bins the most significant parts of the light curve first, to maximise
the time-resolution in those bins. In detail the algorithm is thus:

\begin{enumerate}
\item{Set $n=1$.}
\item{Bin the light curve (or light-curve sections) into new bins of
$n$ original 4-ms bins. If any 4-ms bins are left over at the end of the
light curve (or section), append them to the last new bin in the curve
(or section).}
\item{Search for any of these new bins with SNR above the threshold.}
\item{Save any above-threshold new bins in the output light curve.}
\item{Split the light curve into sections, separated by times which were
saved to the output light curve in the previous step.}
\item{If at least one new above-threshold bin was found during this pass,
increase $n$ by one; otherwise, double $n$.}
\item{If $n$ corresponds to a bin size of $\ge40$\ s simply add every 40-s long bin to the output light
curve.}
\item{If there are still bins in the original light curve which have not
been assigned to the output light curve, go back to step 2. Otherwise,
save the output light curve.}
\end{enumerate}

This results in a light curve in which bins are either above some SNR
threshold, or 40 s in duration. This upper limit on bin size is set to
preserve some information about the behaviour of the light curve during
low-flux periods. Any bins which had SNR$<3$ (regardless of the input
SNR threshold) are flagged as unreliable, and not be shown in the online
plots but they are included in the online data files (on lines beginning
with an `!' -- the comment-line delimiter for {\sc qdp}). Using this above
algorithm, light curves with SNR thresholds of 4,5,6 and 7 are created. 

The BAT-XRT combined light curve will necessarily be presented in log-log
space most of the time, thus data at $t\le0$ are not shown. Although one
could simply use the light curve just created and ignore the bins before
the trigger time, this will result in sub-optimal binning of the light
curve. Instead a second set of light curves are built using the above
algorithm, but given an input 4-ms-binned light curve which only
contains data after the trigger time.

As well as the SNR-binned light curve, constant time-binned 15--150 keV
light curves are also created, using {\sc batbinevt}. These are
nominally created for bin sizes of 4 ms, 64 ms, 1 s and 10 s, but if $\tn<10 $\ s then
the light curves with bin size longer than \tn\ will not be built.
Also, if $\tn>10 $\ s the 4-ms binned light curve is not built, since
this rapidly becomes extremely large. These light curves are
accumulated only over the time range:

\begin{equation}
T_{90,{\rm start}}-1.05\tn \le t \le T_{90,{\rm end}}+1.05\tn
\end{equation}

\noindent where $T_{90,{\rm start}}$ and $T_{90,{\rm end}}$ mark the time range
over which \tn\ was measured by {\sc batgrbproduct}. This is to avoid
filling the light curve with `empty' bins from times where the GRB was
inactive. 

A BAT hardness ratio defined as (25--50 keV)/(15--25 keV)  is also
created using the SNR-binning algorithm above, except that the SNR in
\emph{both\/} bands must be above the threshold before an output bin is
created. The input 4-ms light curves in these bands are created with
{\sc batbinevt}. By default a hardness ratio with a SNR threshold of 5
is created. If this fails to produce at least two bins with
SNR\,$\ge$\,3 the threshold is reduced to 4, and if necessary further to
3. Even if there are still fewer than three bins with an SNR\,$\ge$\,3
this hardness ratio will be used. The errors on the hardness ratio are
calculated assuming that the individual bands' errors are Gaussian.

\subsection{Counts-to-flux conversion factors}
\label{sec:cf}

In order to convert the count-rate light curves (described above) into
flux light curves, a time-evolving counts-to-flux conversion factor is
needed. We do not have sufficient photons to create spectra with a
reasonable time-resolution, so instead we assume a spectral shape and
use the hardness ratios to track the evolution. GRB spectra tend to be
power-laws, or broken power-laws. For the BAT data is has been found
that sometimes a cut-off power-law best describes the data (Sakamoto
\etal2008).

We therefore fit the BAT \tn\ spectrum with both power-law and cut-off
power-law spectral models. If the latter gives a \chisq\ value at least nine
lower than the former (i.e.\ a 3-$\sigma$ improvement) then a cut-off
power-law model is used in all of the counts-to-flux conversion factor
calculations (for BAT and XRT). This occurs for 15 per cent of GRBs; for
the rest a power-law model is used. Note that using a single hardness
ratio, we cannot track both the spectral index and the cut-off energy,
but we lack the statistics to use multiple hardness ratios; we therefore
keep the cut-off energy fixed at the value determined in the above fit,
allowing no evolution of this parameter. This may introduce a systematic
error or bias in our counts-to-flux conversion factors (hereafter `CF'),
which is discussed in Section~\ref{sec:cpl}.

In addition to the emission model, an absorption model must also be
determined. Following the XRT spectrum repository (Evans \etal2009), we
take the absorption model to consist of a {\sc phabs} component with the
Galactic column density (from Kalberla \etal2005) and a second {\sc
phabs} component to represent the absorption local to the burst. If the
redshift of the GRB is in the public domain, this component is a {\sc
zphabs} with the redshift of the GRB.

For GRBs with XRT data we can use the XRT data to determine the values
of these absorption components. Butler \&\ Kocevski (2007a) note that
absorption derived from early-time XRT data, when strong spectral
evolution is present, can be misleading. Therefore, if there are at
least 200 X-ray photons detected at $T>4000$ s post-trigger, a new
spectrum is built using only the data from this time (using the software
presented in Evans \etal2009). If there are fewer than 200 late-time
photons, the absorption values are taken from the XRT spectrum on the
repository\footnote{http://www.swift.ac.uk/xrt\_spectra}. If there are
no XRT data the Galactic absorption component is taken from Kalberla
\etal(2005), and the intrinsic component is assumed to have $N_H=10^{22}$
\cms\ and lie at $z=2.23$; these being the mean values from the XRT
catalogue (Evans \etal2009). The absorption values used, and their
provenance, are given on the Burst Analyser web page for each GRB.

The spectral model thus determined is loaded into {\sc xspec} and the
photon index of the (cut-off) power-law, \G, is stepped from $-1$ to
$5$ in steps of $0.1$. For each \G\ value the hardness ratio, unabsorbed
0.3--10 keV flux, unabsorbed 15--50 keV flux, model normalisation and
count-rate predicted by the model are recorded\footnote{The count-rate
is the observed, i.e.\ absorbed count rate.}. This gives a look-up
table of hardness ratio versus conversion factors (and \G\ values). The
hardness ratios and count-rates are determined for the bands used in
Section~\ref{sec:cr}, i.e. for BAT the hardness ratio is (25--50
keV)/(15--25 keV) and the count-rate is determined over the 15--150 keV
range. For XRT the hardness ratio is (1.5--10 keV/0.3--1.5 keV). The
normalisation of the power-law and cut-off power-law models in {\sc
xspec} is defined as the flux density at 1 keV, in units of photons
keV$^{-1}$ \cms\ s$^{-1}$. A normalisation of one is thus equivalent to
662.5 $\mu$Jy. This can then be extrapolated to give the flux density at
10 keV using either:

\begin{equation}
F_{10 {\rm keV}} = F_{1 {\rm keV}} \times \left(\frac{10}{1}\right)^{-(\G-1)}
\end{equation}

\noindent for the power-law model, or
\begin{equation}
F_{10 {\rm keV}} = F_{1 {\rm keV}} \times \left(\frac{10}{1}\right)^{-(\G-1)} {\rm e}^{-10/E_c}
\end{equation}

\noindent for the cut-off power-law, where $E_c$ is the cut-off energy.

Given these look-up tables, the hardness ratios created in
Section~\ref{sec:cr} can be converted instead into time evolution
histories of conversion factors. For each bin in the original hardness
ratio, the conversion factors and \G\ value appropriate to that hardness
are determined by interpolating within this lookup table. The
uncertainties in the hardness are also propagated by interpolating the
1-$\sigma$ limits on the hardness ratios. For BAT data, because the bins
may not exceed 40-s in duration, it is possible for some bins to have
negative hardness ratios (from Poisson fluctuations of low significance
bins). It is impossible for the lookup table to contain negative values,
so these bins are skipped (and will be interpolated across in
Section~\ref{sec:fluxlc}). Some bins may have positive hardness ratio
values, but negative lower limits. In this case the error-bar on the
conversion factor (or \G) will be truncated at a hardness ratio tending
to zero. See Section~\ref{sec:errors} for more information.

In order to make the (linear) interpolation as accurate as possible it is
preferable to perform it in a phase-space that gives an approximately
linear relationship between hardness ratio and the conversion factor.
By inspecting look-up tables we thus chose the phase-spaces
given in Table~\ref{tab:interp} for the interpolation.

For some GRBs, there were too few photons for even a single hardness
ratio bin to be created for XRT. This is most commonly the case for
bursts detected by missions other than \Swift, which are often not observed by
XRT until many hours after the trigger. In these cases  \G\ and the
conversion factors were determined from the same spectrum used to obtain
the absorption details (above); this does not allow for spectral
evolution (see Section~\ref{sec:sameGamma}).

\begin{table*}
\caption{The phase spaces used to interpolate from hardness ratio (HR) to counts-to-flux
conversion factors (CF) and spectral photon index (\G).}
\label{tab:interp}
\centering
\begin{tabular}{ccccc}
\hline
Instrument        &  0.3--10 keV flux    &  15--50 keV flux  & 10 keV Flux density  & \G \\
\hline
BAT               & log(HR), log(CF)      &  HR, CF          & log(HR), log(CF)     & log(HR), \G \\
XRT               & HR, CF                &  log(HR), log(CF)& log(HR), log(CF)     & log(HR), \G \\

\hline
\end{tabular}
\end{table*}

\subsection{Flux light curves}
\label{sec:fluxlc}

For each bin in each count-rate light curve created in
Section~\ref{sec:cr}, a counts-to-flux conversion factor is determined
by interpolating the time evolution histories of conversion factors created in
Section~\ref{sec:cf}. For BAT this interpolation is done in linear time
space, for XRT log(time) space (since BAT includes negative times and a
small time range, whereas XRT only has times after the trigger and
covers many decades); the conversion factor is interpolated in log space
for both instruments since it covers several decades. The count-rate and
error are then multiplied by the conversion factor to give the flux
for each bin. For BAT data, some hardness ratio bins with negative values
of the ratio may have been skipped (Section~\ref{sec:cf}) in which case
the conversion factor is simply interpolated across this gap. Because
GRBs show strong spectral evolution at early times, it is unwise to
extrapolate the hardness ratio beyond times where we have a reliable
measurement. Thus any light curve bins that occur after the end of the
final valid (i.e.\  positive-valued) hardness ratio bin are discarded
and not converted into flux. At later times the spectral evolution is
minimal (Butler \& Kocevski 2007b), thus for XRT any light
curve bins which occur after the final hardness ratio bin are converted
to flux using the conversion factors from the final hardness bin.

The uncertainty in the conversion factor is not propagated into the
flux light curve. This is because it is in part a systematic effect
and it will dilute the significance of genuine variability in the light
curve. However, we require this information to determine whether
features in the flux light curves are believable or not. Therefore, in
addition to the basic flux light curve plots, we produce plots with
subpanels showing either the conversion factor and its error, or the
derived photon index (\G) and its errors. Users can use these to
consider how much weight should be applied to any given features.

As noted in Section~\ref{sec:cr}, for BAT SNR-binned light curves the
maximum permitted bin duration is 40 s, which can result in some low
significance (SNR $< 3$) bins. These are not shown in the online plots,
however they appear in the data files as lines beginning with an
exclamation mark.

For BAT we also create a flux light curve with no spectral evolution
(for XRT this is already provided by the Light Curve Repository). The
creation is analogous to the light curves described above, but
instead of using the hardness ratio to determine the conversion factor,
a single conversion is used for all bins. This is taken from the
model fitted to the \tn\ spectrum. There are thus no sub-plots available
for the non-evolving spectrum, instead the spectral model and conversion
factors used are given on the Burst Analyser results page for each GRB.

\section{Caveats and checks}
\label{sec:caveats}

Although substantial efforts have been made, by human analysis of the
data products, to ensure that these products are reliable and accurate, 
there are some assumptions inherent to the light-curve creation process
which may introduce artefacts. These are both predictable and
identifiable.  Several of the issues concern the uncertainty involved in
extrapolating the BAT or XRT data beyond the range covered by the
instrument; if there is any reason to question the assumptions involved
in the light-curve creation process, we recommend that users consider
the data in their `own' bands (i.e. the BAT data in the 15--50 keV range
and the XRT data in the 0.3--10 keV range). For the combined BAT-XRT
light curve the `safest' data to use is the flux density at 10 keV, as,
for this dataset, the spectral extrapolation is minimal, as is the
associated uncertainty.

In order to determine whether the data are reliable or potentially prone
to inaccuracies, there are five questions that should be asked of a
dataset.

\subsection{Is \G\ discontinuous between BAT and XRT?}
\label{sec:badgamma}

\begin{figure}
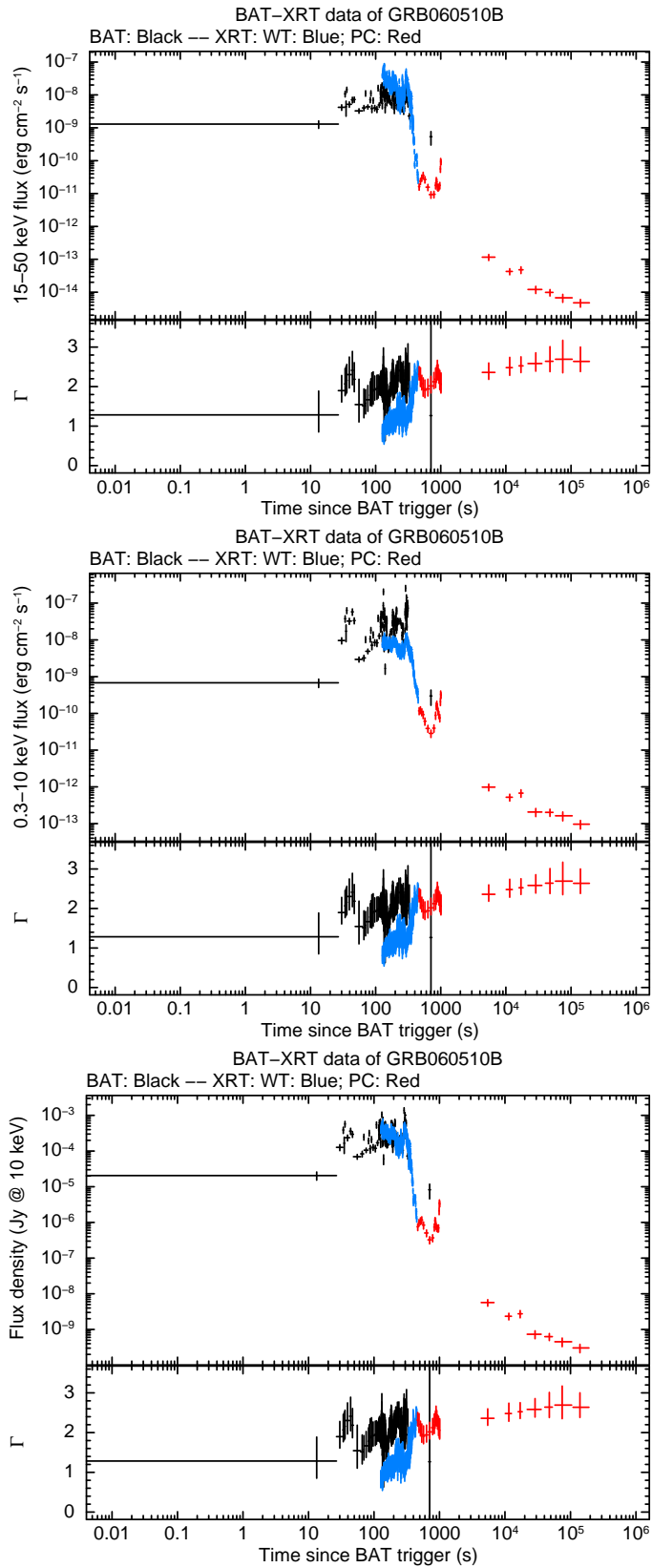

\resizebox{\hsize}{!}{\includegraphics[angle=-90,width=8.0cm]{fig4a.eps}}
\resizebox{\hsize}{!}{\includegraphics[angle=-90,width=8.0cm]{fig4b.eps}}
\resizebox{\hsize}{!}{\includegraphics[angle=-90,width=8.0cm]{fig4c.eps}}
\caption{Combined BAT-XRT light curve of GRB 060510B. The \G\ panel shows
that the BAT and XRT data are not joined by a single power-law
spectrum, as is assumed when extrapolating the flux outside of the
instruments' bandpasses. In the 15--50 keV band light curve (top panel) the XRT flux is
overestimated compared to the BAT flux, and the inverse is true in the
0.3--10 keV light curve (middle panel). However as the lower panel
shows, the 10 keV flux density light curve is much less vulnerable to
this problem.}
\label{fig:badgamma}
\end{figure}

An integral part of the flux conversion is the assumption of a spectral
shape (Section~\ref{sec:cf}). This can be a power-law or cut-off
power-law, but in each case when the flux is extrapolated outside an
instrument's bandpass (i.e. beyond the band over which the shape was
determined) there is an implicit assumption that the power-law index
remains the same. The temporal evolution of the photon index (\G) as
determined from the hardness ratios can be seen as a sub-plot on the
light curves; the default display of the web pages shows this sub-plot.
If the BAT and XRT data show discontinuous \G\ values then the spectral
shape changes between bands and the BAT flux extrapolated to the XRT
band will be unreliable, and vice-versa. This discontinuity is present
in about 20 per cent of cases. Although the flux density is
not immune to this effect, is will be less strongly affected since it
lies close to both bandpasses. This is demonstrated in
Fig.~\ref{fig:badgamma}.

\subsection{Is the BAT spectral fit a cut-off power-law?}
\label{sec:cpl}

\begin{figure}
\resizebox{\hsize}{!}{\includegraphics[width=8.0cm]{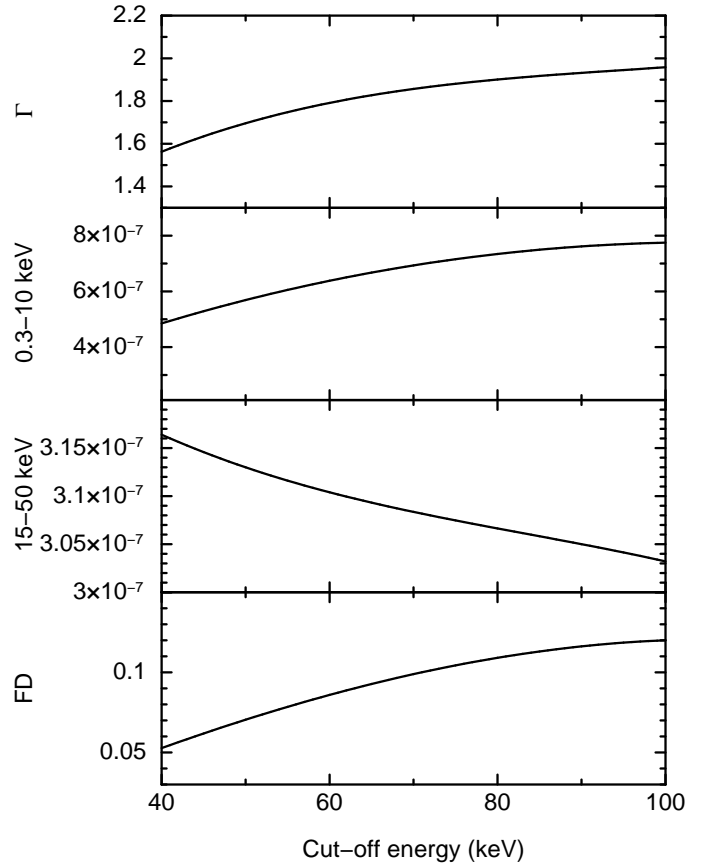}}
\caption{The effect of the cut-off energy on the conversion factors and
\G\ value determined for a given hardness ratio. Panels (top to bottom)
are the spectral index, counts-to-0.3--10-keV flux conversion factor, 
counts-to-15--150-keV flux conversion factor and counts-to-0.3--10-keV
flux density conversion factor. See Section~\ref{sec:cpl} for details.}
\label{fig:cpl}
\end{figure}

As noted in Section~\ref{sec:cf}, in some cases a cut-off power-law
gives a better spectral fit than a power-law (in the 465 BAT light
curves created for the Burst Analyser up to GRB 100316D, 67 required
cut-off power-laws),  however in these cases we do not allow the
high-energy cut-off to vary. If the  cut-off energy moves to lower
energies with time, then at later times the flux above the cut-off will
be overestimated; the inverse is true if the cut-off energy evolves to
harder energies. To investigate the magnitude of this problem we created
a series of BAT spectra for a range of \G\ and cut-off energy ($E_c$)
values. For each spectrum we determined the hardness ratio and
count-to-flux conversion factors as in Section~\ref{sec:cf}. Then, for
each spectrum with a hardness ratio in the range 0.9--0.92 (i.e.\
spectra of approximately equal hardness), we show in Fig.~\ref{fig:cpl}
how the conversion factor and \G\ necessary to generate such a hardness
ratio vary with $E_c$. As can be seen, the $E_c$ dependence is  not
large (note the linear axes). For example, if $E_c$ was frozen at 80
keV, but in reality at late times it was 60 keV, the conversion factors
would be inaccurate by up to 5 per cent. For more extreme examples of
$E_c$ variation (e.g. from 80 keV to 40 keV) these inaccuracies can
reach 50 per cent; however only when extrapolating outside of the BAT
band. The conversion from BAT counts to 15--50 keV flux is never off by
more than 2 per cent.

The range of cut-off energies is limited to 40--100 keV in
Fig.~\ref{fig:cpl} because at higher energies the 15--50 keV spectrum
(whence the hardness ratio is obtained) is insensitive to the cut-off
energy, and with lower values the hard band has so few counts that 
we cannot obtain a hardness ratio. For the XRT we do not see evidence for
the cut-off power-law spectra, so freezing the cut-off energy should not
affect the XRT flux conversions; except that if the cut-off energy were
actually evolving through the 15--50 keV region the use of a fixed
energy would lead to the XRT 15--50 keV flux values being overestimates.

In summary, for the \til15 per cent of GRBs where a cut-off power-law is
preferred to a power-law to fit the BAT spectrum, the flux extrapolated
outside of the instruments' bandpasses may be subject to inaccuracies
of a factor of at most two.

\subsection{Are the \G\ values outside the range 0--4?}
\label{sec:oddGamma}

Although there is considerable scatter in the \G\ values found from BAT
and XRT data, values outside the range \til0--4 are not generally seen
(see, for example, the Swift Data
Table\footnote{http://swift.gsfc.nasa.gov/docs/swift/archive/grb\_table/};
the BAT catalogue, Sakamoto \etal2008; the XRT catalogue, Evans
\etal2009); thus if the \G\ subplot shows values outside this range it
may indicate a low SNR hardness ratio point rather than a real value of
\G\ and users should double-check the size of the errors on \G\ and the
conversion factor. Recall that the errors on the conversion factor are
not propagated into the flux errors (see Section~\ref{sec:errors} for
more). Most light curves show one or two bins with these `extreme'
values; usually they are the last few bins in that BAT light curve and
it is safe to disregard them.

Note also that the hardness-ratio-to-conversion-factor look-up table is only created
for \G\ values in the range $-1$ to 5; if a data point has a value outside
of this range then the conversion factor has been extrapolated rather
than interpolated, and may be less reliable.

\subsection{Are the errors on \G\ or the conversion factor large or
asymmetric?}
\label{sec:errors}

As discussed in Section~\ref{sec:fluxlc}, the uncertainty on the
hardness ratio is propagated to the uncertainty on the conversion
factor, but not into the final error-bar on the flux values. This is
because, where the spectral evolution is not large, these errors are
systematic in nature and tend to wash out genuine variability. However,
if the errors on the conversion factor (visible in the sub-plots) are
large compared to the actual flux values, users should not ascribe
weight to apparently discrepant points. For example if a single data
point lies significantly above the rest of the light curve and has a
conversion factor which is inconsistent with the other conversion factor
values and has a large error, the apparently discrepant flux is almost certainly
the result of a poorly constrained spectral index, whose error is not
reflected in the final light curve.

Because the hardness ratio errors were determined assuming Gaussian
statistics in each BAT band, low SNR hardness bins can have negative
error-bars which extend below zero. However, the
hardness-ratio-to-conversion-factor look up table cannot contain
negative hardness ratio values. Thus any datapoints corresponding to
hardness ratios consistent with zero (i.e. non-detections in one of the
bands) will have uncertainties on their \G\ and conversion factor values
which are truncated where the hardness ratio goes to zero. This can
usually be spotted by a strongly asymmetric error on \G\ or the
conversion factor. Further, the hardness ratio and its corresponding \G\ and
conversion factors is viewable online and users can easily check if and
when it becomes unreliable. There is no straightforward means of
determining what the uncertainty in conversion factor should be at such
times; it should instead be considered unconstrained and users are
advised to ignore these datapoints. As with `extreme' values of \G\
(Section~\ref{sec:oddGamma}), most light curves contain a few bins which
are subject to this issue, typically at the end of the BAT data.

\subsection{Do most/all of the bins for BAT or XRT have the same \G\ or
conversion factor value?}
\label{sec:sameGamma}

For low SNR datasets, there may be only one or two hardness ratio
points. In this case the \G\ subplots will show very little evolution.
This is the case for \til3 per cent of BAT light curves and \til 20 per
cent of XRT light curves -- many of the latter are GRBs detected by other
missions which \Swift\ did not observe until several hours after the
trigger. For BAT-triggered GRBs, only 13 per cent of XRT hardness ratios
have fewer than three bins. This is not a problem, it simply highlights
a lack of information due to the low significance of the data, but users
should note in this case that the light curves tend towards the
non-evolving ones; we lack the data quality necessary to track the
spectral evolution.

\section{Data availability and creation policy}
\label{sec:avail}

The data are all available online from:

\begin{center}{\bf http://www.swift.ac.uk/burst\_analyser/} \end{center}

The top-level page provides various means of choosing a specific burst,
alternatively the trigger number or target ID of the object (if known)
can be appended to the above URL.

The light curves exist both as plots (in postscript and png [Portable
Network Graphics] format) and as text files. For each GRB there are up
to four categories of plot available: BAT-XRT combined data (only
includes data after the trigger time), BAT-only data, XRT-only data, and
a BAT light curve with no spectral evolution. For bursts where BAT or
XRT data do not exist, not all of the above will be created. As
discussed in Sections~\ref{sec:cr} and~\ref{sec:fluxlc} there are
several different BAT-binning criteria used, and the light curves may
have no sub-plot, or a panel showing the conversion factor or \G\ value
for each bin. Each light curve is also created in three different bands:
the 0.3--10 keV flux, 15--50 keV flux, and flux density at 10 keV. The
web page by default presents a single light curve (flux density, with a
\G\ sub-plot) for each of the categories, however users can change which
plots are shown. Also available to download are the conversion factor
time evolution histories, and a tar archive containing all of the data
for the GRB in question. Full documentation supporting these web pages
and downloadable files is also provided online at:

\noindent http://www.swift.ac.uk/burst\_analyser/docs.php

\noindent This page also contains a list of any changes made after
publication of this paper.

The light curves are automatically created and updated via {\sc cron}
jobs which run every ten minutes. These check both BAT and XRT data to
determine whether products need to be built or updated. For BAT data the
products will be built for the first time once data appear on on the
quick-look site \footnote{http://www.swift.ac.uk/access/ql.php}
(typically a few hours after the trigger), and updated when the {\sc
ontime} keyword of these data increases. For XRT data the products will
be built or updated whenever the X-ray light curve from the Light Curve
Repository$^{\ref{foot:lc}}$ is created or
updated.

Users can determine when a product was last updated from the web page
for the GRB. At the top of the page, beneath the GRB name, are given
details of the last dataset which was used to create the product. Also,
in the footer bar at the base of the page, the time (in UT) when the
page was last created is also specified.

\section{Usage}
\label{sec:usage}

The Burst Analyser data products are publically available and may be
freely used. Users should consider the caveats in this paper
(Section~\ref{sec:caveats}), and online (through the documentation link,
above) before using the Burst Analyser data in any scientific medium.

Wherever these data products are used we ask that this paper be cited.
The suggested wording is: ``For details of how these light curves were
produced, see Evans \etal(2010).''

Please also include the following paragraph in the Acknowledgements
section: ``This work made use of data supplied by the UK Swift Science
Data Centre at the University of Leicester.''

\section{Acknowledgements}
This work made use of data supplied by the UK Swift Science Data Centre at the
University of Leicester. PAE, JPO, KLP, APB CP and RLCS acknowledge STFC
support, DNB, acknowledges support from NASA contract NAS5-00136.

\end{document}